\documentstyle[preprint,aps]{revtex}
\begin{document}
\def \beq{\begin{equation}}
\def \eeq{\end{equation}}
\def \bea{\begin{eqnarray}}
\def \eea{\end{eqnarray}}
\title{Coulomb Drag at the Onset of Anderson Insulators}
\author{Efrat Shimshoni}
\address{Department of Mathematics-Physics, Oranim--Haifa University,
Tivon 36006, Israel.}
\date{\today}
\maketitle
\begin{abstract}
I study the Coulomb drag between two identical layers in the Anderson 
insulting state. The dependence of the trans--resistance $\rho_t$ on the
localization length $\xi$ is considered, for both a Mott insulator and
an Efros--Shklovskii (ES) insulator. In the former, $\rho_t$ is
monotonically increasing with $\xi$; in the latter, the presence of a Coulomb
gap leads to an opposite result: $\rho_t$ is {\it enhanced} with a 
{\it decreasing} $\xi$, with the same exponential factor as the single layer
resistivity. This distinction reflects the relatively pronounced role of
excited density fluctuations in the ES state, implied by
the enhancement in the rate of hopping processes at low frequencies.
The magnitude of drag is estimated for typical
experimental parameters in the different cases. It is concluded that a
measurement of drag can be used to distinguish between interacting and
non--interacting insulating state.
\end{abstract}


\narrowtext
\section{Introduction and Principal Results}
\label{sec:intro}
A rich variety of phenomena in disordered electronic systems is associated
with the Anderson localization\cite{localization}. The fundamental effect is a
manifestation of quantum coherence: 
a subtle destructive interference of multiple partial
waves in a random medium leads to exponentially localized electron eigenstates.
In three-dimensional (3D) conductors, a metal--insulator transition (MIT)
can be driven by tunning either the disorder potential or the carrier density
across a threshold value. The transition is identified as second order:
it is associated with a divergence of a correlation length $\xi$, which in
the insulating side has the physical significance of a localization length.
A true phase--transition of this kind is generically absent in
two-dimensions (2D), at vanishing magnetic fields\cite{scaloc}: 
instead, a 2D system exhibits a smooth
cross--over between a ``weak insulator" (in which at finite temperatures $T$ 
the conductivity $\sigma$ acquires a negative logarithmic
correction), and a ``strong insulator" (characterized by an exponentially small
$\sigma(T)$). The presence of a magnetic field $B$, as well as 
spin--orbit scattering (\cite{scaloc},\cite{MA}), 
dramatically alter this behavior - they suppress
localization, and in principle can recover the metallic state in 2D. Most
prominently, in strong magnetic fields, singular extended states play an
important role in the mechanism for the quantum Hall (QH)
effect\cite{QHE}. Recent experimental studies have posed new challenges
in the understanding of Anderson insulators. For example, in QH
systems, the observation of a weak--to--strong localization cross--over
near filling fraction $1/2$, is possibly an interesting indication 
for $B=0$ localization of ``composite Fermions"\cite{CFloc}. 
In contrast, a true MIT has been observed at $B=0$\cite{krav}, 
whose origin is yet obscure. In all the above, electron--electron 
interactions and their interplay with the disorder\cite{el-el}
play an important role, and complicates the theoretical analysis. 

In view of the ongoing research activities described above, it is
highly desired to have an extended array of different probing
techniques. In the present paper, I suggest that Coulomb drag in a double
layer system is potentially an interesting probe, that can diagnose
subtle differences between distinct insulating states.

Coulomb drag\cite{REF:Pog} is a manifestation of the coupling between two
spatially separated systems of charge carriers, due to Coulomb
interactions across a barrier separating them. In the presence of
a transport current in one layer, density fluctuations in that layer
exert a frictional force on the other, and consequently induce a voltage
in the latter (in an open circuit configuration) -- even when tunneling
between the layers is suppressed. The strength of this effect is
characterized by the trans--resistivity, $\rho_t\equiv E_2/j_1$, where $E_2$
is the parallel electric field induced in layer 2
in response to a current density $j_1$ established in layer 1.
Experimentally, drag has been observed in various semiconductor 
heterostructures\cite{dragexp}; theoretically, it has been a subject 
of much recent activity \cite{dragt1,ZM,ESSS,dragt2,BFHJ,nu12,VM}. 
In particular, it was suggested as a useful test of certain electronic
states in the layers, such as compressible states in QH
systems\cite{ESSS,nu12}, and superfluid electron--hole condensates\cite{VM}.

As pointed out in Ref. \cite{ESSS}, inter--layer drag can probe the dynamics
and response properties of electronic systems in a domain that is inaccessible
to transport measurements in a single layer. As it stems from a
frictional force, Coulomb drag is dominated by the interaction between 
relatively long--lived density fluctuations in the two layers.
Therefore, its low temperature behavior is sensitive to the ability of
the electronic systems to create and maintain
such fluctuations, reflected by the density--density response functions
$\chi_{1,2}(q,\omega)$ at {\it low frequencies} ($\omega\rightarrow 0$)
and {\it finite wave--vectors}\cite{isotq} $q$. To see this, note that
to leading order in the screened inter--layer
interaction $U$, the trans--resistivity can be expressed as \cite{ZM}
\begin{equation}
\rho_t={\beta\hbar^2\over \pi n^{(1)}
n^{(2)}e^2}\int{d^2 q \over(2\pi)^2}q^2\int_0^\infty d\omega
{|U(q,\omega)|^2{\rm Im}\chi_1(q,\omega)
{\rm Im}\chi_2(q,\omega)\over 4{\rm sinh}^2(\beta\hbar\omega/2)}\; ;
\label{Rt12}
\end{equation}
here $\beta=1/k_BT$ where $T$ is the temperature, $n^{(1)},n^{(2)}$
are the carrier densities in layers 1,2, and
\beq
U(q,\omega)={V(q)e^{-qd}\over
[1+V(q)\chi_1(q,\omega)][1+V(q)\chi_2(q,\omega)]
-V^2(q)e^{-2qd}\chi_1(q,\omega)\chi_2(q,\omega)}
\label{Uqw}
\eeq
where $V(q)=2\pi e^2/\epsilon q$ ($\epsilon$ being the background dielectric
constant), and $d$ the inter--layer distance. In Eq. 
(\ref{Rt12}), the integration over $\omega$ is effectively cutoff by $T$, 
while the cutoff on $q$ is set by $1/d$; the former is {\it stricter} at 
low $T$.

In the vicinity of a MIT, the behavior of drag is
expected to be particularly interesting, due to the significant role played by
density fluctuations. Variations in the conductivity (of one layer or
both) have competing effects on the {\it availability} of density
fluctuations and on their 
{\it stability}. In the conducting state (e.g., in the diffusive regime
\cite{ZM}), the drag is {\it enhanced} as the diffusion coefficient is reduced,
since then the decay rate of density fluctuations is slowed down. On the other
hand, in the insulating state the creation of density fluctuations is 
suppressed at low $q$ and $\omega$, and this can lead to a {\it reduction} of 
$\rho_t$ with a reduced localization length -- provided that the
simultaneous variations in the screening properties are not dominant
\cite{BFHJ,screen}. Thus, in contrast with the single
layer resistivity, $\rho_t$ is potentially a non--monotonous function of the
parameter which drives the transition; a similar behavior is expected
also near a smooth (non00critical) cross--over between a weak and strong
insulator. At the transition, density fluctuations
and their dynamics become critical, and one expects a pronounced peak in the
drag (similarly to the behavior predicted in \cite{ESSS} for QH transitions).

In this paper I study the Coulomb drag in Anderson insulators, in order
to verify under what circumstences the above described qualitative picture
holds. I calculate $\rho_t$ in a double--layer system, where the electronic
states in the layers are assumed for simplicity to be identical and 
uncorrelated. Two possible insulating states are considered: a Mott 
insulator\cite{Mott}, where the in--layer long--range Coulomb interactions 
are suppressed, and an Efros--Shklovskii (ES) state\cite{ES} 
where the interactions are significant. In both cases, the dominant 
transport mechanism is assumed to be variable range hopping\cite{hopping} 
among localized sites, and $\rho_t$ is evaluated as a function of the 
localization length $\xi$.

The resulting trans--resistance indicates a dramatic difference between
the Mott and ES insulators. While the former exhibits a suppression of
$\rho_t$ with a decreasing $\xi$, as implied by the naive argument
pointed out earlier, in the latter the dominant contribution to
$\rho_t$ is {\it enhanced} with the same exponential factor as the
single layer resistivity. As will be shown below, this follows 
from the relative enhancement of hopping processes at $\omega\rightarrow 0$,
due to the presence of a Coulomb gap in the ES state. Drag
measurement is therefore a suggestive experimental means of
distinguishing the two types of insulating states. In contrast, the
single layer resistivity exhibits a qualitatively similar dependence on
$\xi$ and $T$: $\rho(T)\sim e^{(T_0/T)^\alpha}$, where in the 2D Mott
insulator $\alpha=1/3$ and $T_0\sim\xi^{-2}$, and in the ES state 
$\alpha=1/2$ and $T_0\sim\xi^{-1}$.

In the following sections I detail the calculation of $\rho_t$: in
Sec. \ref{sec:Mott}, for the Mott insulator, and in Sec. \ref{sec:ES} 
for the ES insulator. The conclusions and experimental implications 
are summarized in Sec.  \ref{sec:sum}.

\section{Mott Insulators}
\label{sec:Mott}
I consider two parallel 2D layers separated by a perfect insulating barrier of
thickness $d$, in both of which the electronic state is a non--interacting
Mott insulator. The trans--resistance is evaluated using Eqs.
(\ref{Rt12}), (\ref{Uqw}), where $\chi_1(q,\omega)=\chi_2(q,\omega)$ is
the density response function, which has been derived diagramatically
by Vollhardt and W{\"o}lfle \cite{VW}. To leading order in $q$ and
$\omega$, it is given by 
\beq
\chi_{1(2)}(q,\omega)={dn\over d\mu}{Dq^2\over
Dq^2-[i\omega+\tau(\omega^2-\omega^2_0)]}\; .
\label{chiVW}
\eeq
Here $dn/d\mu$ is the density of states, $\tau$ is the ellastic mean free
time, and $D=v_F^2\tau/2$ ($v_F$ being the Fermi velocity) is the 
diffusion coefficient in the conducting state. The localization effect
is represented by the frequency $\omega_0$: as pointed out in Ref. \cite{VW},
Eq. (\ref{chiVW}) is consistent with a simple hydrodynamical model, in which
localization is manifested as an effective restoring force acting on density
fluctuations, with a characteristic oscillator frequency $\omega_0$.
The localization length is directly related to $\omega_0$ through 
$\xi=v_F/\sqrt{2}\omega_0$.

I next assume that $T$, the effective upper cutoff on $\omega$ in
Eq. (\ref{Rt12}), is sufficiently low that the 
$\omega\ll\omega_0,\omega_0^2\tau$.
The real and imaginary parts of $\chi(q,\omega)$ (hereon, the layers indices
are dropped) are then given, respectively, by
\bea
{\rm Re}\chi(q,\omega)\approx {dn\over d\mu}{q^2\over q^2+\xi^{-2}}\, ,
\label{Rchi}
\\
{\rm Im}\chi(q,\omega)\approx {dn\over d\mu}{Dq^2\omega\over
D^2(q^2+\xi^{-2})^2}\; .
\label{Ichi}
\eea
Substitution in Eqs. (\ref{Rt12}) and (\ref{Uqw}) yields the final
expression for $\rho_t$. The result depends crucially on whether $\xi$
is smaller or greater then the layers separation $d$, and exhibits a
cross--over between the two limiting cases considered below.

Deep in the insulating state, $\xi\ll d$, one can neglect the $q^2$
term in the denominators of Eqs. (\ref{Rchi}) and (\ref{Ichi})
(in the effective range
of wave--vectors, $q<1/d$ and hence $q\ll 1/\xi$). The inter--layer 
interaction can be then approximated by its unscreened form,
$U(q,\omega)\approx V(q)e^{-qd}$. The integrations in Eq. (\ref{Rt12})
are streight--forward and result in
\beq
\rho_t\approx {5\over 32\pi}{h\over e^2}\left({k_BT\over \hbar
Dn}\right)^2(q_{TF}d)^2\left({\xi\over d}\right)^8\; ,
\label{RtMs}
\eeq
where $n\equiv n^{(1)}=n^{(2)}$, and $q_{TF}=2\pi
e^2(dn/d\mu)/\epsilon$ is the Thomas--Fermi screening wave--vector.

When the localization length is increased (e.g. by controlling a
parameter that drives the insulator into a transition to the metallic
state), eventually the distance $d$ is exceeded. In the limit $\xi\gg
d$, the screening of $U(q,\omega)$ becomes significant as long as 
$q>1/\xi^2q_{TF}$ (see Eqs. (\ref{Uqw}) and (\ref{Rchi})). 
The integration over $q$ in (\ref{Rt12})
is therefore divided into two regions: for $0<q<1/\xi^2q_{TF}$,
$U(q,\omega)$ is approximated by the unscreened form, while for
$q>1/\xi^2q_{TF}$,
\beq
U(q,\omega)\approx{\pi e^2q[1+1/(\xi q)^2]^2\over
\epsilon q_{TF}{\rm sinh}(qd)}\; .
\label{UMw}
\eeq
The dominant contribution to $\rho_t$ in this regime is 
\beq
\rho_t\approx {1\over 48\pi}{h\over e^2}\left({k_BT\over 
\hbar Dn}\right)^2{{\rm ln}(q_{TF}\xi)\over(q_{TF}d)^2}\; .
\label{RtMw}
\eeq
Note that in both Eqs. (\ref{RtMs}) and (\ref{RtMw}) $\rho_t$ is an
increasing function of $\xi$ (typically $q_{TF}\sim 1/d$, 
hence $q_{TF}\xi\gg 1$). However, for $\xi\gg d$ the dependence on
$\xi$ is weaker: there is a considerable contribution to the frictional drag
from density fluctuations on length scales within $\xi$. In addition, in
this regime the enhancement of screening competes with the enhancement
of conductivity.

Finally, when $\xi$ is further increased, eventually the approximations made in
the beginning of this section break down. For $\xi\rightarrow\infty$,
the form of $\chi(q,\omega)$ coincides with that of the diffusive
state considered in Ref. \cite{ZM}. In the low $T$ limit, where
$k_BT\ll\hbar/\tau$, Zheng and MacDonald find $\rho_t\sim T^2{\rm
ln}(k_BT\tau/\hbar)/D^2$. Notice that the $T$--dependence is altered in the 
opposite limit $k_BT\gg\hbar/\tau$: then, in the relevant range of
$q$ and $\omega$, ${\rm Im}\chi(q,\omega)\sim (dn/d\mu)Dq^2/\omega$, and
the trans--resistance is 
\beq
\rho_t\approx {1\over 64\pi^3}{h\over e^2}{1\over (q_{TF}d)^2(n\ell^2)} 
\left({k_BT\over \hbar Dn}\right)
\label{Rtdiff}
\eeq
where $\ell\equiv v_F\tau$ is the mean free path.
When a control parameter is varried so that $\xi$ gradually increases
from $\xi\ll d$ to $\xi\rightarrow\infty$, $\rho_t$ interpolates
between the expressions (\ref{RtMs}), (\ref{RtMw}) and  (\ref{Rtdiff}).

\section{Efros--Shklovskii Insulators}
\label{sec:ES}
The effect of Coulomb interactions within the layers on the dissipative 
processes associated with the hopping mechanism is quite subtle 
\cite{ES}. The dominant processes, at small but finite $\omega$,
$q$ and $T$, are transitions of electrons between two localized sites that
are close in energy\cite{ESbook,Efros}; typically, different pairs of such 
sites are sparsely distributed. In the presence of Coulomb interactions, 
on one hand the transitions are enhanced due to the greater
probability to find a {\it singly} occupied pair; on the other hand, a
Coulomb gap is introduced in the single--electron density of states
near the Fermi level. The {\it a.c.} conductivity $\sigma(q,\omega)$,
assisted by resonant transitions, is more strongly affected by the
former, and is therefore enhanced compared to the non--interacting
case. This has crucial consequences on the drag between layers at the
ES insulating state.

Similarly to the previous section, the trans--resistance is evaluated
employing Eqs. (\ref{Rt12}), (\ref{Uqw}) with the appropriate form of 
$\chi(q,\omega)$, assumed to be identical in the two layers. In this
case the effect of $\chi(q,\omega)$ on the inter--layer interaction can
be neglected, and $U$ is approximated by the unscreened form,
$U(q,\omega)\approx V(q)e^{-qd}$. As will become evident below, the
$\omega$--integration in Eq. (\ref{Rt12}) has an infra--red divergence,
and hence the prominent contribution to $\rho_t$ arises from the lower
cutoff. This cutoff is introduced at finite $T$
by dephasing, associated with phonon--mediated
relaxational processes, which suppresses resonant transitions between 
sites separated by a distance $r_\omega$ larger than the dephasing length 
$L_\phi$. The pair arm $r_\omega$ diverges \cite{ESbook} with the frequency as
\beq
r_\omega=\xi{\rm ln}(\omega_0/\omega)\; ,\quad \omega_0\equiv
{e^2\over\epsilon\xi\hbar}\; ,
\label{romega}
\eeq
and $L_\phi$ is set by the hopping distance
\beq
L_\phi=\xi\left({T_0\over T}\right)^{1/2}\; ,\quad
\left(T_0={\hbar\omega_0\over k_B}\right)\; .
\label{lphi}
\eeq
Hence, coherent frequency--driven hopping occur at $\omega>\omega_c$ where
\beq
\omega_c=\omega_0e^{-L_\phi/\xi}=\omega_0e^{-(T_0/T)^{1/2}}\; .
\label{omegac}
\eeq

To proceed with the calculation of $\rho_t$ using Eq. (\ref{Rt12}), I relate
${\rm Im}\chi(q,\omega)$ to the a.c. conductivity through
\beq
{\rm Im}\chi(q,\omega)={q^2\over \omega e^2}\sigma(q,\omega)\; .
\label{IchiSig}
\eeq
$\sigma(q,\omega)$ at a finite $q$ was calculated for a 2D ES insulator by 
Aleiner and Shklovskii \cite{AlSh}, who obtained
\bea
\sigma(q,\omega)\sim {\cal C}_1\epsilon\xi\omega\quad{\rm for}\quad 
q\ll r_\omega^{-1}\; ,\nonumber\\
\sigma(q,\omega)\sim {{\cal C}_2\epsilon\xi\omega\over (q r_\omega)^2}
\quad{\rm for}\quad r_\omega^{-1}\ll q\ll \xi^{-1}\; ,\label{ASsig}\\
\sigma(q,\omega)\sim {{\cal C}_3\epsilon\xi\omega\over (q r_\omega)^2
(q\xi)^{2-\eta}}\quad{\rm for}\quad \xi^{-1}\ll q\; ;
\nonumber
\eea
here ${\cal C}_j$ ($j=1,2,3$) are numerical constants of order unity, and
$\eta$ is non--vanishing in case the localized single electron states have 
multi--fractal structure. Note that Ref. \cite{AlSh} considers the limit
$\hbar\omega\gg k_B T$, while in the present case the relevant range of 
frequencies (which dominates Eq. (\ref{Rt12})) is $\hbar\omega<k_B T$; 
however, following Refs. \cite{Efros} and \cite{PSh}, it can be shown that the
result differs only by the values of the  ${\cal C}_j$'s. Inserting 
Eq. (\ref{ASsig}) in  (\ref{IchiSig}) yields approximate expressions for 
${\rm Im}\chi(q,\omega)$ in three different regimes of $q$.
I then evaluate $\rho_t$, similarly to the previous section, distinguishing the
limit cases $\xi\ll d$ and $\xi\gg d$. 

In the case $\xi\ll d$, the high $q$ regime ($q>\xi^{-1}$) is 
exponentially suppressed, and the $q$--integration in Eq. (\ref{Rt12})) 
yields (for carrier density $n$ in the two layers)
\beq
\rho_t\sim {{\cal C}_2^2\beta\hbar^2\over 8n^2e^2}\left({\xi\over d}\right)^2
\int_{\omega_c}^{\omega_d}{d\omega\over 
{\rm sinh}^2(\beta\hbar\omega/2)r_\omega^4}\; .
\eeq
The upper limit $\omega_d\equiv\omega_0 e^{-2d/\xi}$ corresponds to 
$r_\omega=2d$; the integration over the frequency range $\omega>\omega_d$ 
gives a subdominant contribution, that is neglected here. The final expression
for $\rho_t$ is dominated by the lower cutoff (for sufficiently low $T$, 
$L_\phi\gg d$):
\beq
\rho_t\sim {{\cal C}_2^2 \over 4\pi}{h\over e^2}{1\over (nd\xi)^2}
\left({T\over T_0}\right)^3
\exp\left\{\left({T_0\over T}\right)^{1/2}\right\}\; .
\label{RtESs}
\eeq

In the opposite regime, where the localization length $\xi$
greatly exceeds $d$, one should
account for the contribution of short length--scale density
fluctuations with $1/\xi\ll q\ll 1/d$. The integration over $q$ is
facilitated by the approximation $r_\omega\gg d$ (note that in the
relevant renge of $\omega$, $r_\omega>\xi$), and gives
\beq
\rho_t\sim {\beta\hbar^2\xi^2\over 2n^2e^2}
\int_{\omega_c}^{\omega_d}{d\omega\over
{\rm sinh}^2(\beta\hbar\omega/2)}\left\{
{({\cal C}_1^2/ 6-{\cal C}_2^2)\over r_\omega^6}+
{({\cal C}_2^2+{\cal C}_3^2/2(1-\eta))\over r_\omega^4\xi^2}
\right\}\; .
\eeq
Similarly to the short $\xi$ limit, the lower cuttoff dominates.
The final expression for $\rho_t$ for $\xi\gg d$ is
\beq
\rho_t\sim {({\cal C}_2^2+{\cal C}_3^2/2(1-\eta))\over \pi}
{h\over e^2}{1\over n^2\xi^4}
\left({T\over T_0}\right)^3
\exp\left\{\left({T_0\over T}\right)^{1/2}\right\}\; .
\label{RtESw}
\eeq
This result essentially differs from Eq. (\ref{RtESs}) only by the
algebraic dependence on $\xi$ -- short length--scale fluctuations are 
effectively cuttoff by $\xi$ rather than $d$ (the latter sets a higher
upper cutoff on $q$). Hence, there is no dependence 
on the inter--layer separation in this case.

\section{Discussion and Summary}
\label{sec:sum}
As shown in the calculations detailed above, Coulomb drag between
layers in the Anderson insulator state can serve as a clear signature
of the presence or absence of a Coulomb gap in the layers. Most
importantly, the trans--resistance in a double--layer of Mott
insulators is {\it suppressed} with a decreasing temperature and 
localization length; in contrast, the presence of a Coulomb gap in ES
insulating layers leads to a {\it divergence} of $\rho_t$ at low $T$
and $\xi$. 

The physical origin of this distinction is the relative enhancement
of resonant hopping processes in the presence of a Coulomb gap, which
ensures a greater probability that a ``destination site'' is empty. 
In the single--layer
a.c. conductivity $\sigma(q,\omega)$, this difference is indicated by the
power--law dependence on $\omega$ ($\omega$ versus $\omega^2$), though in
both cases $\sigma$ vanishes at $\omega\rightarrow 0$. However, the low
frequency limit of the dynamical structure factor, 
\beq
S(q,\omega)={\hbar\over 1-e^{-\hbar\omega\beta}}
{\rm Im}\chi(q,\omega)\; ,
\eeq
is crucially distinct: since $S\sim \sigma/\omega^2$,
in the Mott insulator $S(q,\omega\rightarrow 0)$ is finite, while in
the ES state it diverges as $1/\omega$. As expressed by Eq.
(\ref{Rt12}), the frictional drag directly probes the dynamics of
density fluctuations through the convolution of
$S(q,\omega)$ in the two layers. Therefore, in the ES insulator
$\rho_t$ depends on the {\it lower} frequency cutoff, associated with the
dephasing length, hence diverges at low $T$ and $\xi$. In the Mott
insulator this anomaly does not exist, and $\rho_t$ decreases with $\xi$
due to the suppression of excited density fluctuations -- in agreement
with the naive intuitive argument.

In both types of insulators, $\rho_t(\xi)$ depends on whether $\xi$ is
smaller or larger than the layers separation $d$. The predictions of this
paper can be summarized as follows: (a) in the Mott insulator, $\rho_t\sim
T^2$ similarly to the free electron gas case. The localization is
manifested as a strong dependence on $\xi$ for $\xi\ll d$,
\beq
\rho_t\propto (\xi/d)^8\; ,
\eeq
while for $\xi\gg d$ the subtle role played by screening effects yields
a weaker dependence,
\beq
\rho_t\propto {\rm ln}(q_{TF}\xi)\; .
\eeq
(b) in the ES insulator, the resulting $\rho_t(\xi, T)$ is
\bea
\rho_t\propto {1\over (\xi d)^2}\left({T\over T_0}\right)^3
\exp\left\{\left({T_0\over T}\right)^{1/2}\right\}\quad(\xi\ll d)\\
\rho_t\propto {1\over \xi^4}\left({T\over T_0}\right)^3
\exp\left\{\left({T_0\over T}\right)^{1/2}\right\}\quad(\xi\gg d)\; ,
\eea
where $k_BT_0=e^2/\epsilon\xi$. Note that $\rho_t$ diverges at low $T$
and $\xi$ with the same exponential factor as the in--layer resistance
$\rho$. The algebraic prefacor typically suppresses $\rho_t$ with
respect to $\rho$ by 3 to 4 orders of magnitude. This ensures that although
the drag is a huge effect in this case, the two--layer resistivity
tensor is still almost diagonal, and the weak--coupling assumption
underlying Eq. (\ref{Rt12}) is justified.

Experimentally, a crossover from an ES to a Mott insulator can be in
principle controlled by a metallic back gate, which effectively
attenuates the range of interactions in the layer. The above predictions
imply that at the ES state, the onset of an insulating behavior should
be accompanied by a sharp increase of $\rho_t$ towards the insulating
regime; when the insulating state is well approximated by the Mott behavior,
$\rho_t$ should be peaked near the transition to the insulator, and
strongly attenuated when $\xi$ is reduced below $d$. A similar
qualitative behavior is also expected in case where some of the
simplifying assumptions of this paper are violated -- for example, when
the layers are not identical, and in particular if only one of them
undergoes a transition to the insulator. However, note that in the case
where one of the layers is a good conductor, it may serve as a back
gate which suppresses Coulomb interactions in the other. 

In order to observe an appreciable drag at low $T$ in the Mott state, 
the desired experimental setup should involve low mobility, low density
samples. For $\ell\sim 1\,\mu m$, $n\sim 10^{10}\,cm^{-2}$,
$d=200\,{\AA}$, band mass of $m\approx 0.07m_e$, $\epsilon=13$
(typical to GaAs) and $T=1\,K$, I get $\rho_t$ in the order of a few
tens of $m\Omega$ in the regime $\xi>d$. When the localization length
is reduced to $20\,{\AA}$, $\rho_t$ is attenuated by a factor of
$10^{-5}$. Assuming the same parameters in the ES state, I estimate
$\rho_t\sim 575\,\Omega$ for $\xi\sim 1000\,{\AA}$; at $\xi\sim 100\,{\AA}$,
the trans--resistance rises to $\rho_t\sim 100\,k\Omega$.

Finally, the experimental testing of the effects predicted in this paper 
involves a number of difficulties. Primarily, once the in--layer resistance
is comparable to that of the barrier separating the two layers,
tunneling across the barrier is no longer negligible; its contribution should
be carefully eliminated. In addition, to obtain a sizable signal, the voltage
imposed on the drive layer in an insulating state may be large enough
to produce non--linear response effects. The role of inter--layer coupling
mediated by phonons (see, e.g., Gramila {\it et al.} in Ref.
\cite{dragexp}, and Ref. \cite{BFHJ}), and possible thermoelectric
effects (Solomon  {\it et al.} in \cite{dragexp}, Laikhtman {\it et al.}
in \cite{dragt1}), are not discussed in the present paper. These are
generally expected to be subdominant at low $T$; however, at the onset
of an insulating state (and particularly near a MIT) such effects may
be enhanced as well. Nevertheless, the estimates made above indicate that
Coulomb drag near the onset of an insulator is an appreciable effect, 
and a sensitive probe of the significance of Coulomb interactions
in insulating states.

\acknowledgements

I gratefully acknowledge useful discussions with J.\ Eisenstein, M. Fogler,
T. Gramila, A. MacDonald, B. Shklovskii and S. Sondhi, and particularly with
U. Sivan who also informed me of preliminary experimental results.
This work was supported by the Technion -- Haifa University 
Collaborative Research Foundation.   

\end{document}